# Analytical determination of the reach-through breakdown voltage of bipolar transistors, asymmetric thyristors and Punch Through-IGBTs

Miron J. Cristea
'POLITEHNICA University of Bucharest', Bucharest, Romania
E-mail: mcris@lydo.org  Tel: 40-214-024-915  Fax: 40-214-300-555

## *Introduction*

In this work, an analytical formula that gives the reach-through breakdown voltage of bipolar transistors, asymmetric thyristors, Punch Through-IGBTs and other devices with similar structure is deduced for the first time.

## *Breakdown voltage determination*

In an earlier work [1] we have established the formula:

$$\int_{SCR} \frac{x}{\varepsilon} \rho(x) dx = V_{bi} - V_F = V_{bi} + V_R \quad (1)$$

where $\varepsilon$ is the dielectric permittivity of semiconductor, $\rho(x)$ - the density of electric charge in the space charge region of the junction (SCR), $V_{bi}$ is the built-in potential of the junction, $V_F$ - the external forward bias applied on the junction and $V_R$ - the reverse voltage applied to the junction - if it is reverse biased.

This formula will be used to determine the breakdown voltage by the reach-through of the base of a bipolar transistor (Fig.1). The semiconductor is supposed to be homogenous (e.g. silicon).

$$-\frac{q}{\varepsilon}\int_{-W_p}^{0} x\left[N_0 \exp(-\frac{x^2}{L_d^2}) - N_B\right]dx + \frac{q}{\varepsilon}\int_{0}^{W_n} xN_B dx = V_{biC} + V_R \quad (2)$$

In equation (2) $q$ is the elementary electric charge, $N_0$ is the surface concentration of the base diffusion, $N_B$ is the constant concentration of the $n^-$ central layer, $L_d$ the technological diffusion length of the base doping and $V_{biC}$ is the built-in potential of the collector junction. $W_n$ is the extension of the SCR in the lightly doped $n^-$ layer, equal with the physical $n^-$ layer width because of the $n$ field stopper layer and $W_P$ is the extension of the depletion region in the base (Fig.1.):

$$\frac{q}{\varepsilon}\left[\frac{L_d^2}{2}N_0\exp\left(-\frac{x^2}{L_d^2}\right)+\frac{x^2}{2}N_B\right]\bigg|_{-W_P}^{0}+\frac{q}{2\varepsilon}N_B x^2\bigg|_{0}^{W_n}=$$

$$=\frac{qL_d^2}{2\varepsilon}N_0\left[1-\exp\left(-\frac{W_P^2}{L_d^2}\right)\right]-\frac{q}{2\varepsilon}N_B W_P^2+\frac{q}{2\varepsilon}N_B W_n^2=$$

$$=V_R+V_{biC} \tag{3}$$

At reach-through:

$$\frac{qL_d^2}{2\varepsilon}N_0\left[1-\exp\left(-\frac{W_{pRT}^2}{L_d^2}\right)\right]-\frac{q}{2\varepsilon}N_B W_{pRT}^2+\frac{q}{2\varepsilon}N_B W_n^2=$$

$$=V_{BRT}+V_{biC} \tag{4}$$

where $V_{BRT}$ is the reach-through breakdown voltage and $W_{pRT}$ is the depletion region in the base at reach-through equal to the physical width of the base - $W_{B0}$, less the space charge region width of the p-side of the emitter junction, $W_{SEP}$. (We suppose that the transistor does not break before reach-through, by attaining the critical field at the collector junction.)

$$W_{pRT}=W_{B0}-W_{SEP} \tag{5}$$

with $W_{SEP}$ - the depletion region extended in the base from the emitter [2]:

$$W_{SEP} = L_d \left\{ \sqrt{-\ln\left[e^{-\frac{x_{je}^2}{L_d^2}} - \frac{2\varepsilon}{qN_0 L_d^2}(V_{biE} - V_{BE})\right]} \right\} - x_{jE} \qquad (6)$$

where $x_{jE}$ is the emitter-base metallurgical junction depth (Fig.1.), $V_{biE}$ is the built-in potential of the emitter junction and $V_{BE}$ - the external bias applied on the emitter-base junction.

The resulting formula of the reach-through breakdown voltage is:

$$V_{BRT} = \frac{qN_D}{2\varepsilon} \cdot (W_n^2 - W_{pRT}^2) + \frac{qL_d^2 N_0}{2\varepsilon}\left(1 - e^{-\frac{W_{pRT}^2}{L_p^2}}\right) - V_{biC} \qquad (7)$$

In the case of asymmetric thyristors and the PT-IGBTs (Punch Through IGBTs), the value of the reach-through voltage of the p-base is the same. The structural difference is the presence of a $p^+$ layer at the exterior (Fig.1.) that does not influence the reach-through of the base at the p-n⁻ junction.

### *Acknowledgement*

The author thanks Miss Rocsana Ionescu for carefully preparation of the drawing.

### *Conclusion*

In this work, an analytical formula was calculated for the reach-through breakdown voltage of semiconductor devices like bipolar transistors, asymmetric thyristors, Punch Through IGBTs and other devices with similar structure.

### *References*

[1] Miron J. Cristea, **Integral equation for electrically charged space regions - theory and applications**, arXiv:math-ph/0608068, 31 Aug 2006.

[2] Miron J. Cristea, **First time calculation of the depletion region width and barrier capacitance of practical diffused semiconductor junctions**, arXiv:math-ph/0609004, 1 Sep 2006.

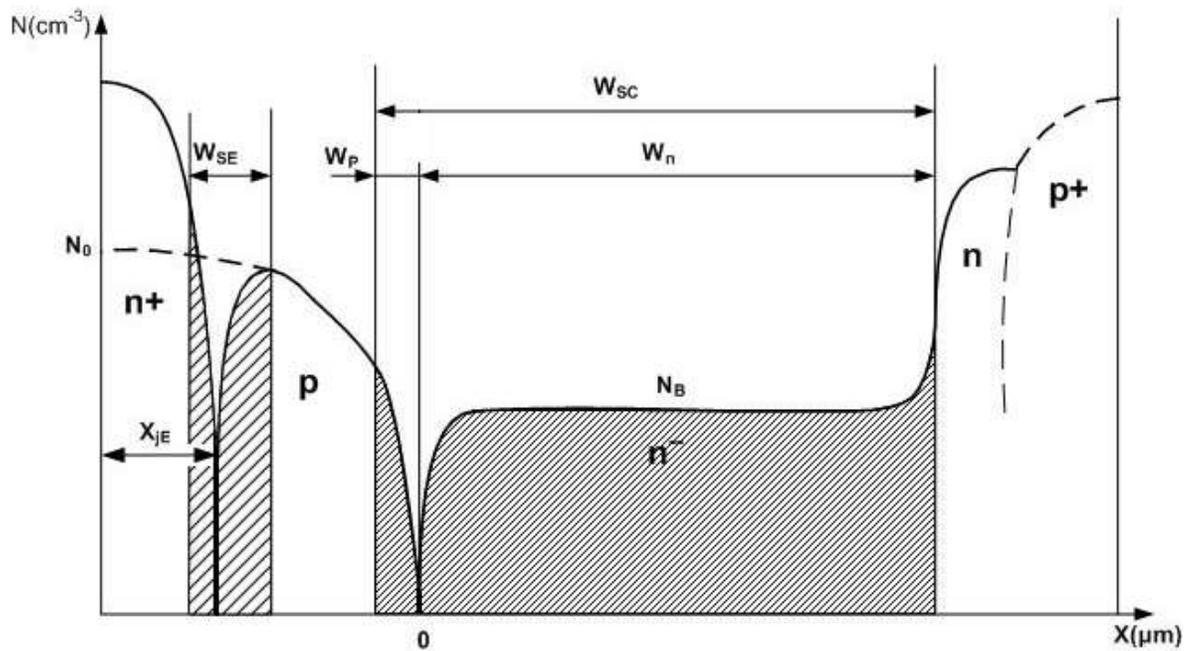

**Figure 1.** The doping profile of a bipolar transistor - without the p+ layer, respectively of an asymmetric thyristor/PT-IGBT - with the p+ layer. The base profile is Gaussian.